\documentclass[]{spie}  

\usepackage{amsmath,amsfonts,amssymb}
\usepackage{graphicx}
\usepackage[colorlinks=true, allcolors=blue]{hyperref}
\usepackage{array}
\usepackage{multirow}

\title{SAMplus: adaptive optics at optical wavelengths for SOAR}

\author[a]{Daniel M. Faes}
\author[b]{Andrei Tokovinin}
\author[a]{Tarcio Vieira}
\author[c]{Alexandre Mello}
\author[a]{Marcia Domingues}
\author[a]{Denis Andrade}
\author[d]{Bruno C. Quint}
\author[e]{Jesulino B. dos Santos}

\affil[a]{Departamento de Astronomia, IAG, Universidade de S\~ao Paulo, S\~ao Paulo, Brazil}
\affil[b]{Cerro Tololo Inter-American Observatory, La Serena, Chile}
\affil[c]{Universidade Tecnol\'ogica Federal do Paran\'a, Curitiba, Brazil}
\affil[d]{SOAR Telescope, La Serena, Chile}
\affil[e]{Laborat\'orio Nacional de Astrof\'isica, Itajub\'a, Brazil}

\authorinfo{Further author information: (Send correspondence to DMF: Departamento de Astronomia IAG/USP -- Rua do Matão, 1226, Cidade Universitária - São Paulo/SP, Brasil - 05508-090)\\E-mail: moser@usp.br, Telephone: +55 11 3091 2705.}

\begin{document} 
\maketitle

\begin{abstract}
Adaptive Optics (AO) is an innovative technique that substantially improves the optical performance of ground-based telescopes. The SOAR Adaptive Module (SAM) is a laser-assisted AO instrument, designed to compensate ground-layer atmospheric turbulence in near-IR and visible wavelengths over a large Field of View. Here we detail our proposal to upgrade SAM, dubbed SAMplus, that is focused on enhancing its performance in visible wavelengths and increasing the instrument reliability. As an illustration, for a seeing of 0.62\,arcsec at 500 nm and a typical turbulence profile, current SAM improves the PSF FWHM to 0.40\,arcsec, and with the upgrade we expect to deliver images with a FWHM of $\approx0.34$\,arcsec --- up to 0.23\,arcsec FWHM PSF under good seeing conditions. Such capabilities will be fully integrated with the latest SAM instruments, putting SOAR in an unique position as observatory facility.
\end{abstract}

\keywords{Astronomical instrumentation, adaptive optics, high angular resolution, GLAO, visible wavelengths}
\section{Introduction}
Adaptive Optics (AO) is a technique that substantially improves the resolution of ground-based telescopes. So far, most of AO technologies were implemented for small on-sky Field of View (FoV) and/or were applied over Infra-Red (IR) wavelengths. Ground-layer adaptive optics, GLAO\cite{2002ESOC...58...11R}, achieves partial correction of seeing in a wide field by compensating only low-altitude turbulence. As high turbulence remains uncorrected, GLAO offers some improvement over natural seeing, rather than diffraction-limited resolution. The resolution gain from GLAO depends on the turbulence profile, not just on the seeing. On nights when the dominant turbulence is in the high atmosphere, GLAO gives no gain.  

The SAM project started as soon as GLAO started to be discussed by the community. Refs.~\citenum{tokovinin_visible-light_2003} and \citenum{tokovinin_ground_2004} contain the first concepts for the system. It was designed as a laser-assisted instrument to compensate ground-layer atmospheric turbulence in near-IR and visible wavelengths over a relative large FoV (3$\times$3\,arcmin$^2$). 

The SAM module, with its laser launch sub-system, was initially described in Ref.~\citenum{tokovinin_sam_2008}. The module was initially set to operate in natural guide star (NGS) mode, using stars brighter than $V=11.5$ with images over 7.5$\times$10.0\,arcsec$^2$\cite{tokovinin_sam_2010}. SAM performance in laser guide star (LGS) mode was reported in Ref.~\citenum{tokovinin_performance_2012} and the most recent description of the system is done in Ref.~\citenum{2016PASP..128l5003T}. 

SAM is in regular science operations since 2014. During this period, SAM covered multiple science cases. Some examples are binarity of dwarf stars\cite{2014AJ....148...72T}, young-stellar objects\cite{2017ApJ...844...47R} and star forming regions\cite{mendes_de_oliveira_first_2017}; stellar variability\cite{2016AJ....152...55S} and properties and populations of open\cite{2018ApJ...857..132K} and globular\cite{2013AJ....145..165F} clusters, as well as interacting galaxies\cite{murphy_2014}.

This paper describes the project to upgrade SAM, dubbed SAMplus, and is divided as following: section~\ref{sec:samp} describes the project, its goals and management; the expected performance and the parameters of the simulations are in \autoref{sec:simul}. The concepts for the new wave-front sensor (WFS) optics and the mechanical design for the new components are in \autoref{sec:designs}. Our final remarks about the project are in \autoref{sec:conclusions}.

\section{The SAMplus}
\label{sec:samp}

\subsection{Project overview}
SAM was designed in early 2000's, based on the bimorph Deformable Mirrors (DM) technology with 60 actuators and a 10$\times$10 Shack-Hartmann wavefront sensor (WFS) in a 80x80 pixel CCD-39 from e2v. This combination of DM and WFS corrects about 28 strongest optical turbulence modes. Thus, wave-front compensation remains partial. Consequently, the resolution gain offered by SAM strongly depends on the wavelength and seeing. 

At red and near-IR wavelengths and under good seeing, the resolution of SAM approaches the free-atmosphere seeing (i.e. the seeing produced by turbulence above $\sim$1\,km). However, in the $V$ band the resolution gain diminishes. One can say that turbulence correction in SAM is doubly partial: first, not all turbulence is sensed (as inherent to the GLAO concept), and, second, the sensed turbulence is compensated only partially, limited by the correction order. Consequently, SAM's performance could be improved by an higher-order correction of the incoming wavefront, especially at shorter wavelengths. Upgrading key AO components with state-of-the-art technology would considerably enhance the performance of the AO system, particularly the compensation order (smallest size of perturbations that can be corrected).

SAMplus proposes to employ this technology to use the same SAM's Nd:YAG laser flux more efficiently, in combination with a new voice-coil DM with 241 (or 400) actuators and corresponding new WFS and Real-Time Computer (RTC). Details of these components and the expected performance are described in \autoref{sec:simul}.

\subsection{Requirements}
\label{sec:samp.goal}
The SAM AO technology is 20 years old and is largely obsolete. If the DM fails, there would be no replacement. One of the objectives of the project is therefore to reduce the risk of  SAM failure by renovating its key elements.

Another requirement is to maintain compatible optical characteristics of SAM, such as corrected field size, laser operating height, and optics, i.e., an upgrade based on the AO-components only. Also, the changes must be within the instrument budgets of weight, heat dissipation, and volume constraints.

The main objective, however, is to enhance multiple science cases that will benefit from AO corrections at shorter wavelengths. Beyond the obvious areas that are already covered by SAM, such as young stars, star-forming regions, globular clusters and multiple stars, based on imaging (or Speckle techniques), SAMplus will significantly enhance spectroscopic observations. 

Beyond the already available Fabry-Perot interferometer coupled to SAM (and its future upgrade\cite{BTFI2}), it SAM be integrated with IFU spectroscopy\cite{2010SPIE.7739E..4SD} and other SOAR instruments, such as the proposed multi-object spectrograph SAMOS (SOAR Multi-object Spectrograph)\cite{robberto_samos_2016}. Important spectroscopic diagnostic lines lie at visible wavelengths, such as Balmer lines (H$\alpha$ and H$\beta$), forbidden lines [NII], [SII], and [OIII], that will greatly impact studies of compact nebulae, nuclei of galaxies and  gravitational arcs and multiple lensed quasar systems.

The combination of SAMplus with the latest SOAR instrumentation is in strong synergy for follow-up observations of discoveries and transient events that will be detected by the neighboring LSST observatory.

\subsection{Project Management}
SAMplus has Dr. A. Tokovinin as PI, Dr. D. Faes as Co-PI. It is sponsored by FAPESP (The S\~ao Paulo Research Foundation), IAG/USP (Institute of Astronomy and Geophysics of University of S\~ao Paulo) and SOAR direction. The project support is part of the incentives for the development of the group of Astronomical Instrumentation at IAG/USP. Most of workforce, as well as the prototyping activities, will be under the responsibility of the group.

The project is divided in three phases: (i) Conceptual Design; (ii) Critical Design and Laboratory assembly; and (iii) Commissioning at the telescope. The first phase, expected to occur in the second semester of 2018, will be revised by an external committee. The second phase encompasses the procurement of the AO components, and their assembly in the optical laboratory at IAG/USP with mock-up sub-systems simulating the interfaces with the telescope. Once the system is assembled and tested, it will be shipped to SOAR for integration with SAM with the minimum down-time as possible. The project is expected to be completed by early 2021.

\section{Simulations and Expected performance}
\label{sec:simul}
The most important aspect of the SAMplus project is the investigation of two critical AO components: (i) the specifications of the new DM (actuators number or pitch), within a physical size compatible with SAM optics (50\,mm pupil size) and (ii) the definition of the detector camera for the WFS. A detector with substrate gating can eliminate the need of the current Pockels-cell shutter, simplifying the WFS subsystem and increasing the detected flux. The WFS optical strategy tends to be the Shack-Hartmann, since it is a well-established technology. The WFS performance with a pyramid layout will be evaluated, depending on the availability of resources; however, the spot elongation from the low-altitude laser beam shall be a limiting factor for the pyramid.

The laser system will remain the same in SAMplus. So, the same number of photons will be distributed over a larger number of sensing elements to match the increased number of the DM actuators. The optimal value of sensing elements and actuators must be evaluated.

For SAMplus, we are working on three main scenarios: (i) WFS geometry of 16x16 apertures and a corresponding DM geometry of 17x17 actuators (2.95\,mm pitch); (ii) WFS geometry of 19x19 apertures and a corresponding DM geometry of 20x20 actuators (2.50\,mm pitch) and (iii) WFS geometry of 16x16 apertures and a DM geometry of 20x20 actuators. Each WFS sub-aperture can be sampled by 8 pixels or 4 pixels (2 pixels binning) in the camera. The LGS is located at altitude of 7 km, so that the peripheral spots are slightly elongated because of the chosen range gate (0.15\,km, as in the current SAM). 

The atmosphere is modeled by five phase screens. Each screen is characterized by its altitude, fractional strength, and wind speed. The total turbulence strength is defined by the $D/r_0$ ratio at the wavelength $\lambda_{\rm 0} = 0.5\,\mu m$. The turbulence parameters used in the simulations are shown in Table~\ref{table:atturb}. Its also contains the parameters for good, median, and bad atmospheric conditions. The altitudes and wind speeds of the five layers are common for all cases and are based on evaluations of the site made by Ref.~\citenum{Tokovinin2005}.

The simulations presented in this paper were done with the YAO software package\cite{Rigaut2013}. YAO is an open-source adaptive optics Monte-Carlo simulation tool. 

\begin{table}[htbp]
\centering
\caption{Parameters used for atmospheric turbulence simulations.}
\label{table:atturb}
\begin{tabular}{ccccccc}
	\hline
	Layer & 1 & 2 & 3 & 4 & 5 & $D/r_0$ at 0.5\,$\mu m$\\ 
    \hline \hline
    Altitude (km) & 0 & 1 & 2 & 4 & 8 & -\\
	\hline 
	Wind Speed (m/s) & 9 & 10 & 15 & 25 & 25 & -\\ 
	\hline
    Fraction - Good & 0.79 & 0.01 & 0.01 & 0.06 & 0.13 & 24\\
    \hline
    Fraction - Median & 0.74 & 0.02 & 0.02 & 0.10 & 0.12 & 30\\
    \hline
    Fraction - Bad & 0.70 & 0.03 & 0.07 & 0.10 & 0.10 & 37\\
    \hline
\end{tabular}
\end{table}
\par

Important simulation parameters comparing current SAM and SAMplus performance are shown in Table~\ref{table:param}. The wave-front sensor operates at 355~nm wavelength, using as reference a laser beam focused at 7060~m. The system sampling time is of 2~ms. The DM of SAM is a bimorph mirror with concentric rings, and for SAMplus the actuators are in stack-array format. The photon flux for SAM is the actual measured photon flux, divided by two to account for error sources not present in simulation, resulting in realistic values. For SAMplus, a number of possible configurations were considered, and the photon flux is dimensioned accordingly.

\begin{table}[htbp]
\centering
\caption{Parameters used for simulations of SAM and SAMplus}
\label{table:param}
\begin{tabular}{cccc}
	\hline
	Parameter for Simulation & SAM & SAM+ & SAM+\\ 
    \hline \hline
    WFS Subapertures & 10x10 & 16x16 & 19x19\\
    \hline
    Pixels per subaperture & 4x4 & 4x4 or 8x8 & 4x4 or 8x8\\
    \hline
    Pixel size (arcsec) & 0.8 & 1.3 or 0.65 & 1.5 or 0.76\\ 
    \hline
    Photon flux (ke\textsuperscript{-}/sub/s) & 100 & 43 & 30 \\
    \hline
    CCD readout noise (e\textsuperscript{-}) & 3 & 0.3 & 0.3 \\
    \hline
    CCD quantum efficiency & 0.73 & 0.2 & 0.2 \\
    \hline
    DM Actuators & - & 17x17 or 20x20 & 20x20 \\
    \hline
    DM electrodes per ring & 4,8,12,16,20 & - & - \\
    \hline
    Tip-Tilt GS $\lambda$ (nm) & 650 & 650 & 650 \\
    \hline
    Tip-Tilt GS (Vmag) & 16 & 16 & 16 \\
    \hline
    Tip-Tilt GS position (arcmin, off-axis) & 2  & 2 & 2 \\
    \hline
\end{tabular}
\end{table}
\par

The results for these configurations are presented in Table~\ref{table:result05} at the 0.5~$\mu m$ wavelength, in Table~\ref{table:result07} at 0.7~$\mu m$ and in Table~\ref{table:result10} at 1.0~$\mu m$, for the three atmospheric conditions being considered. The values are in terms of Full-Width at Half-Maximum (FWHM, miliarcsec unit), and "No AO" is the SOAR telescope performance without any AO correction.

\begin{table}[htbp]
\centering
\caption{Results in FWHM (mas) at $\lambda=0.5~\mu m$ for the SOAR telescope without AO, using SAM and with SAMplus. The configurations for SAMplus are expressed in terms of number of DM actuadors, WFS subapertures and number of pixels per subaperture.}
\label{table:result05}
\begin{tabular}{ccccccccc}
	\hline
	 & No AO & SAM & SAM+ & SAM+ & SAM+ & SAM+ & SAM+ & SAM+\\\hline
     DM & - & - & 17x17 & 17x17 & 20x20 & 20x20 & 20x20 & 20x20 \\\hline
     WFS & - & 10x10 & 16x16 & 16x16 & 16x16 & 16x16 & 19x19 & 19x19 \\\hline
     px/sub-ap & - & 4x4 & 4x4 & 8x8 & 4x4 & 8x8 & 4x4 & 8x8 \\
    \hline \hline
    Good (mas) & 488 & 283 & 232 & 234 & 238 & 249 & 242 & 243\\
    \hline
    Median (mas) & 618 & 403 & 330 & 345 & 341 & 361 & 342 & 356\\
    \hline
    Bad (mas) & 827 & 554 & 473 & 490 & 500 & 500 & 462 & 493\\
    \hline
\end{tabular}
\end{table}
\par

\begin{table}[htbp]
\centering
\caption{Same as Table~\ref{table:result05}, but for $\lambda=0.7~\mu m$.}
\label{table:result07}
\begin{tabular}{ccccccccc}
	\hline
	 & No AO & SAM & SAM+ & SAM+ & SAM+ & SAM+ & SAM+ & SAM+\\\hline
     DM & - & - & 17x17 & 17x17 & 20x20 & 20x20 & 20x20 & 20x20 \\\hline
     WFS & - & 10x10 & 16x16 & 16x16 & 16x16 & 16x16 & 19x19 & 19x19 \\\hline
     px/sub-ap & - & 4x4 & 4x4 & 8x8 & 4x4 & 8x8 & 4x4 & 8x8 \\
    \hline \hline
    Good (mas)& 440 & 201 & 169 & 183 & 186 & 192 & 174 & 197 \\
    \hline
    Median (mas)& 537 & 320 & 267 & 272 & 262 & 270 & 256 & 276\\
    \hline
    Bad (mas)& 646 & 442 & 365 & 379 & 348 & 361 & 387 & 389\\
    \hline
\end{tabular}
\end{table}

\begin{table}[htbp]
\centering
\caption{Same as Table~\ref{table:result05}, but for $\lambda=1.0~\mu m$.}
\label{table:result10}
\begin{tabular}{ccccccccc}
	\hline
	 & No AO & SAM & SAM+ & SAM+ & SAM+ & SAM+ & SAM+ & SAM+\\\hline
     DM & - & - & 17x17 & 17x17 & 20x20 & 20x20 & 20x20 & 20x20 \\\hline
     WFS & - & 10x10 & 16x16 & 16x16 & 16x16 & 16x16 & 19x19 & 19x19 \\\hline
     px/sub-ap & - & 4x4 & 4x4 & 8x8 & 4x4 & 8x8 & 4x4 & 8x8 \\
    \hline \hline
    Good (mas)& 336 & 166 & 161 & 161 & 161 & 165 & 128 & 147 \\
    \hline
    Median (mas)& 443 & 241 & 228 & 218 & 217 & 215 & 205 & 201 \\
    \hline
    Bad (mas)& 651 & 313 & 285 & 306 & 265 & 290 & 294 & 315 \\
    \hline
\end{tabular}
\end{table}

\begin{figure}[htbp]
   \centering
   \includegraphics[width=0.6\linewidth]{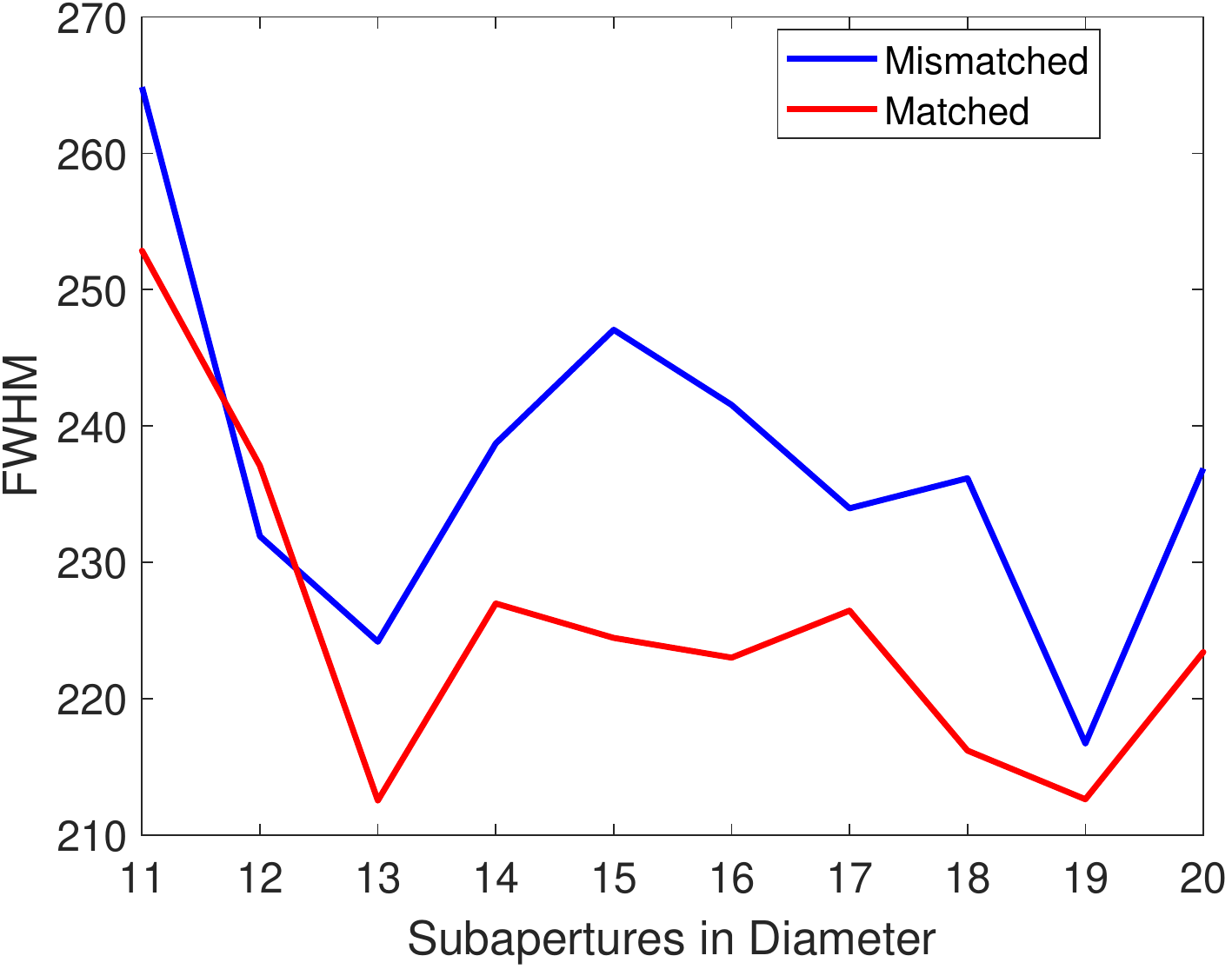}
   \caption{Simulations of the FWHM for SAMplus system matching the number the $N$ WFS subapertures with $N+1$ DM actuators (stack-array size; ``matched'' configuration -- light curve), and varying the $N$ WFS subapertures with a fixed number of 20 DM actuators (``mismatched'' configuration -- dark curve).  Imaging wavelength at $\lambda=0.5\mu m$ and ``Good'' turbulence profile used (Table~\ref{table:atturb}).}
   \label{fig:mismatched}
 \end{figure}

An analysis of the effects of using mismatched number of DM actuators and WFS subapertures was done to help in the deformable mirror definition. This is an idealized environment simulation, where the DM actuators are fixed at 20x20 versus a case where the actuators are matched with the WFS subapertures. For this type of system modal control is needed. The results are shown in Fig.~\ref{fig:mismatched}. The results show a very close response comparing both cases.

The results show considerable improvement from SAM performance and very similar performances for all the different scenarios considered for SAMplus. The degradation of the system with the mismatch number of subapertures and DM actuators is minimal, as the WFS subsystem obtains sufficient photons for the wavefront sampling even in a 19x19 configuration. Once system with more actuators has a potential for better performance, it is preferable the use of the DM with 20x20 actuators.

\section{Conceptual design of SAMplus}
\label{sec:designs}

\subsection{WFS Optical Design}
\label{sec:designs.optics}
The range-gated Rayleigh laser guide star used by SAM requires very short exposure times. As the current E2V CCD-39 does not have a fast readout, the solution used by SAM for fast gating is an electro-optical Pockels cell device. However, these shutters add complexity and lead to losses of light and/or image quality. The Pockels cell is polarization-sensitive and the photon flux in SAM also depends on the rotator angle. By getting rid of the Pockels cell, the WFS optical path will not only be simplified but will also gain in efficiency.

\begin{figure}[htbp]
   \centering
   \includegraphics[width=0.7\linewidth]{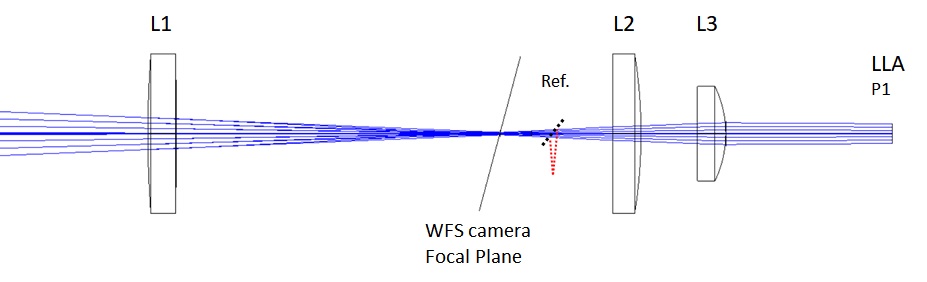}
   \caption{Proposed SAMplus WFS S-H module optical design. }
   \label{opt_design}
 \end{figure}

The SAM WFS path consists of two major subsystems: 1) the ``WFS Camera'' that focuses the collimated beam coming from the DM onto the WFS module's field-stop and 2) the ``Shack-Hartmann WFS Module'' (S-H WFS module) which provides a collimated beam for pupil onto the S-H lenslet array (LLA). The SAMplus will not change the WFS Camera, but the WFS module will be very simplified, as shown Figure.\ref{opt_design}. The new module will be mounted on the WFS focusing stage in the current SAM.

The requirements for this composition is are:

i) Spherical aberration \(<\)100\,nm RMS WF error at LGS 7Km. The altitude at which the spherical aberration of the SAM WFS camera design is the highest.

ii) S-H pupil P1 diameter = 3.16 +/- 0.02mm.

iii) Reasonable space in the region around the WFS Camera focal plane, for mounting a Field-Stop viewing acquisition camera and S-H Reference Light injection.

The optical design is such that it is only necessary to adjust the position of L1 to work with LGS from 7 to 14 km. Table \ref{table:LGS_position} show the relative positions of components for different altitudes.

\begin{table}[htbp]
\centering
\caption{Proposed WFS module configuration to work at different LGS altitudes.}
\label{table:LGS_position}
\begin{tabular}{p{1.5cm}m{1.5cm}m{2.cm}p{2cm}}
\hline
 LGS altitude (Km) & \multicolumn{2}{m{3.3cm}}{WFS Z Position (mm) Fold mirror to} & Relative distance L1-to-Stop \\
\cline{2-3}
 &Lens L1&Field Stop&    \\
\hline\hline
7& 635.5 & 687.1 & 51.6\\
10&555.4&592.9&37.6\\
14&507.4&535.2&27.8\\
\hline
\end{tabular}
\end{table}

The lenses will made of fused Silica with ultra-violet (UV) anti-reflex (AR) coating. The lenses L1 and L2, are 25.4\,mm diameter and L3 is 12.7\,mm diameter.  L2 and L3 re-images the beam on the S-H lenslet array (P1) to a size of 3.16\,+/-\,0.02mm, for all LGS altitudes in the range by only adjusting the position of L1. This value for P1 is valid in the case of the choice for 16x16 subapertures to S-H sensor. The 2-elements design for collimation (L2 + L3) provides adjustment flexibility to the module. The Table \ref{table:Lens_parameter} shows the lenses parameters. The image quality (IQ) of the complete optical train from SOAR entrance pupil to the lenslet array based on optical path differences (OPD), for on-axis and off-axis beam at 2'' field radius, are summarized in Table \ref{table:355aberra}.

\begin{table}[htbp]
\centering
\caption{Proposed WFS S-H module lens parameters.}
\label{table:Lens_parameter}
\begin{tabular}{m{1.5cm}m{1.5cm}m{1.5cm}m{1.5cm}m{1.2cm}c}
\hline
 Element & R1 (mm)  & R2 (mm) & Thinck. (mm) & Material & Comment  \\
\hline\hline
Lens L1 & 169.6 & -1060.4 & 4.5 & Silica & OFR LLU-25-300NUV (UV AR)\\
Lens L2 & 471.01 & -79.99 & 4.5 & Silica & Custom\\
Lens L3 & INF    & -17.53 & 4.5 & Silica & Custom\\
\hline
\\
\end{tabular}
\end{table}

\begin{table}[htbp]
\centering
\caption{Aberrations from LGS (355nm) on-Axis and off-Axis at 2". }
\label{table:355aberra}
\begin{tabular}{c|c|c|c|c|c|c|cc}
\hline
 LGS altitude (Km) & \multicolumn{7}{c}{IQ SOAR +LGS +WFS-camera +LGS-WFS-module (nm WF OPD RMS)}  \\
\cline{2-8}
 &\multicolumn{3}{c|}{On-Axis} & \multicolumn{4}{c}{Off-Axis R\(_{field}\)=2arcsec}   \\
 \cline{2-8}
 & Spherical & Coma & Strehl & Spher. & Coma & Astg. & Strehl\\
\hline\hline
7 & 36.3 & -2.2 & 0.51& 36.3 & 2.8 & 6 &0.6\\
10 & -1.4 & -12.2 & 0.81 & -1.8 &-11.7 & 3.2& 0.6\\
14 & -16.3 & -18.1 & 0.8 &16.6 & -19.9 & 2.13 & 0.73\\
\hline
\multicolumn{8}{c}{} \\
\end{tabular}
\end{table}

SAM contains a built-in artificial UV light source that can mimic focus and pupil of the LGS, and also simulates turbulence. This device, called TurSim, is essential for the SAM and SAMplus operations. The 370\,nm UV light emitted by diode laser in TurSim generates some chromatic aberrations when compared to the LGS light (355\,nm).  As expected, the system have defocus aberration. Another aberrations are listed in Table 8. 

\begin{table}[htbp]
\centering
\caption{Aberration from TURSIM TurSim diode laser (370nm). }
\label{table:370aberra}
\begin{tabular}{c|c|c|c|c|c|c|cc}
\hline
 LGS altitude (Km) & \multicolumn{7}{c}{IQ SAMPlus TurSim (nm WF OPD RMS)}  \\
\cline{2-8}
 &\multicolumn{3}{c|}{On-Axis} & \multicolumn{4}{c}{Off-Axis R\(_{field}\)=2arcsec}   \\
 \cline{2-8}
 & Defocus & Spherical & Coma & Defocus & Spherical & Coma & Astg.\\
\hline\hline
7 & 1.18 & 43.3 & -1.8& 7.77 & 43.3 & -6.6 &-6.3\\
10 & 35.9 & 3.3 & -12.5 & 31.8 &3.4 & -12.6& -7.4\\
14 & 51.8 & -13.3 & -18.1 &59.2 & -13.3 & -16.6 & -8.5\\
\hline
\end{tabular}
\end{table}

The current SAM WFS has 10 x 10 sub-apertures and 8x8 pixels per sub-aperture. It uses the LLA from A$\mu$S with a 0.192\,mm pitch (exactly 8~pixels) and a focal length of 5.75\,mm. Due to high readout noise and to speed up the frame time, the 2x2 CCD binning is currently used, resulting in a FoV per sub-aperture and a pixel size of 3.3$''$ and 0.8$''$, respectively. The proposal for SAMPlus is to keep the LLA pitch and focus, and only increase the number of sub-apertures. If the 16x16 geometry is chosen, the FoV of individual sub-aperture will be 5.2$''$. As in the current SAM, each sub-aperture will use 8x8 pixels in SAMplus, with the new CCD working without binning. The on-sky projected  pixel size will be 0.65$''$. With this value, the atmospheric wavefront sampling is increased, with a one-to-one match of the DM actuators to lenslets apertures.

\subsection{Mechanical Design}
\label{sec:designs.mech}
\begin{figure}[htbp]
   \begin{center}
   \begin{tabular}{cc} 
   \includegraphics[height=0.4\linewidth]{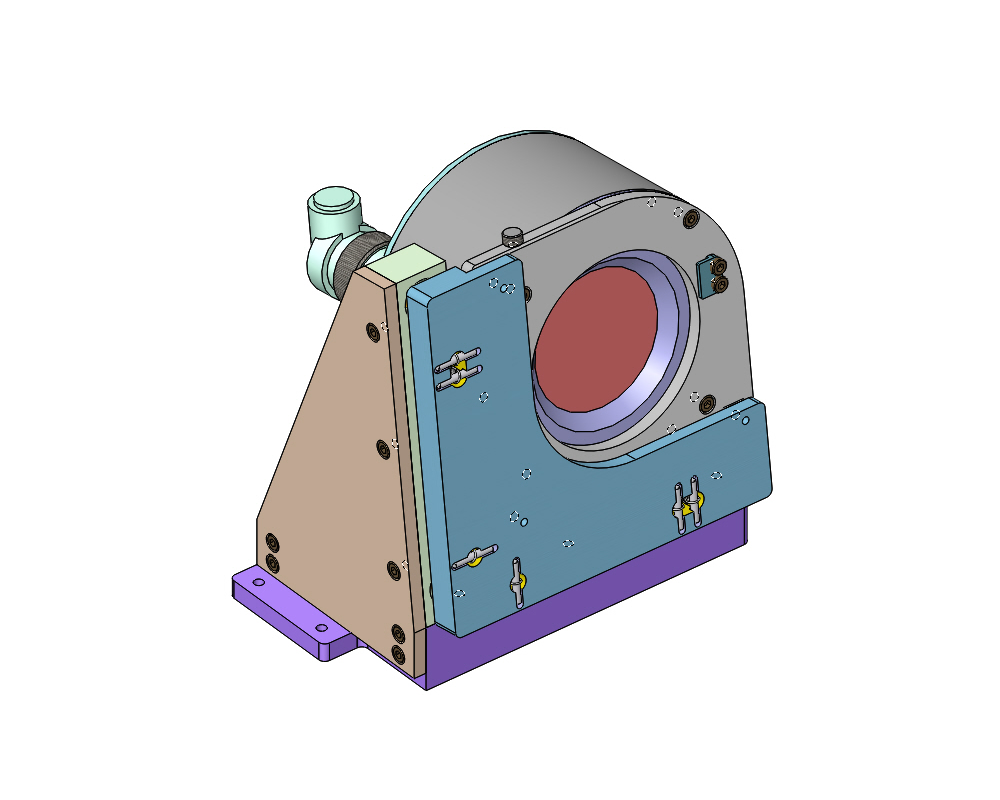} &
   \includegraphics[height=0.4\linewidth]{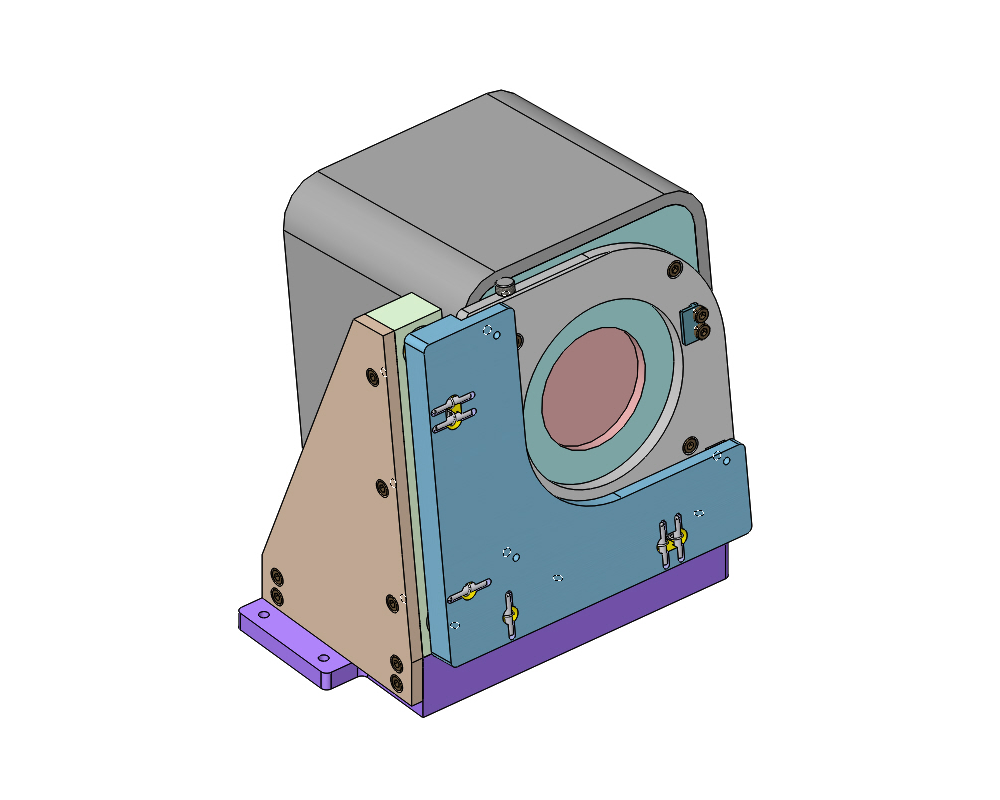} \\
   \end{tabular}
   \end{center}
   \caption[example] 
   { \label{fig:dm} Isometric view of the deformable mirrors (DMs). At left: currently in SAM. At right: proposed one to SAMplus. The main concern is to have enough space inside the instrument for the new DM, as currently its envelope is very close to the WFS camera.}
\end{figure}

The main concern about the new DM is its volume, as SAM is very limited in free space inside the instrument. Currently, the DM is just a few millimeters from the WFS camera. The dimensions for the new DM are in discussions with the manufacturer for the DM mount design. Current and proposed DM mounts are shown in Fig.~\ref{fig:dm}. The expected new DM sizes are around 130 x 150 x 120~mm (width x height x depth), the width being the most restrictive dimension.

The idea is to continue with the fixation of the DM from its the front face if the manufacturer support this option. This allows to use the existing tip-tilt and focus manual adjustment. If not possible, a bracket will be made from the underside of the DM box and a new support for the adjustments will be designed. 

\begin{figure}[htbp]
   \begin{center}
   \begin{tabular}{cc} 
   \includegraphics[height=0.4\linewidth]{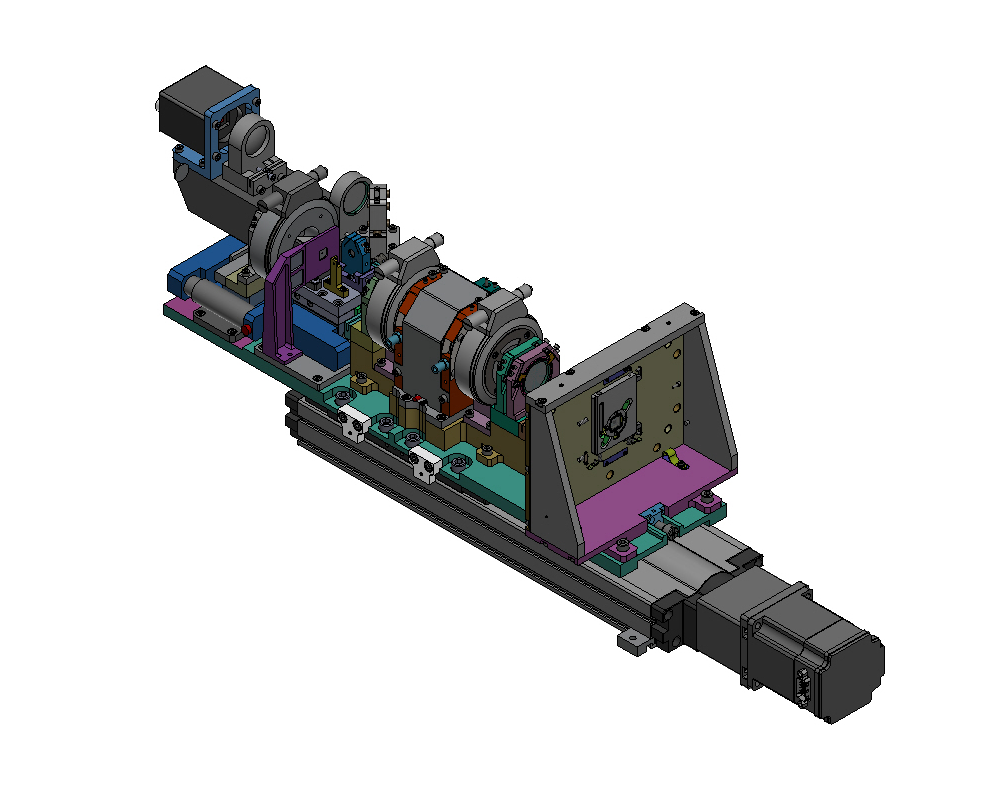} &
   \includegraphics[height=0.4\linewidth]{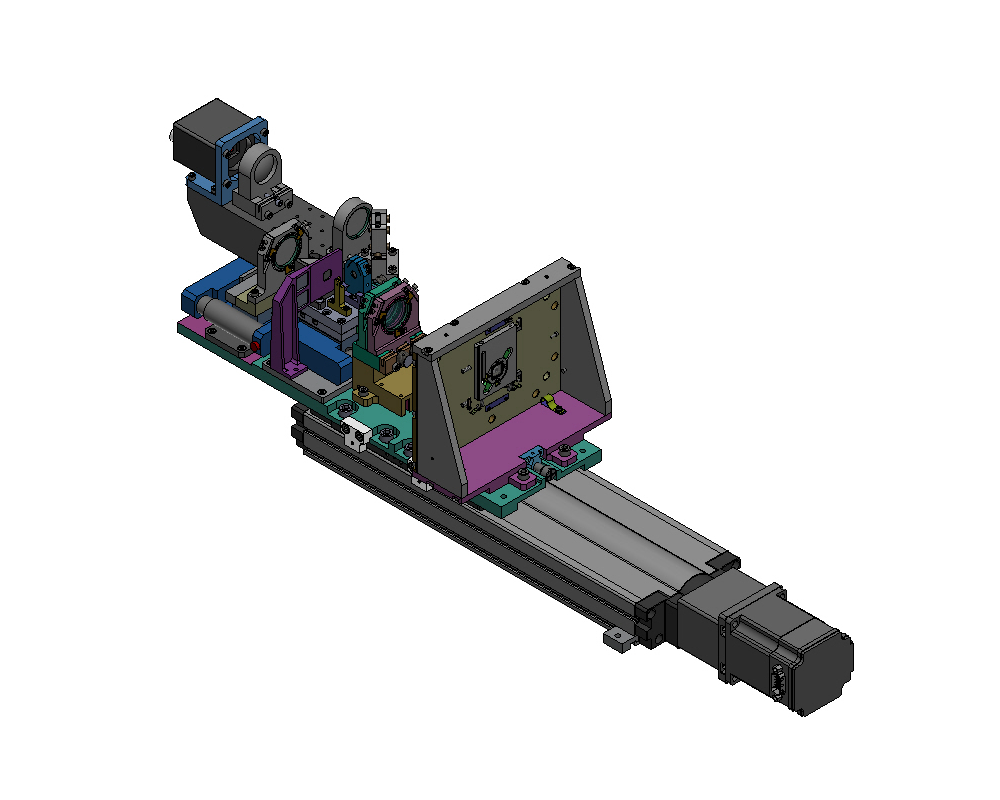} \\
   \end{tabular}
   \end{center}
   \caption[example] 
   { \label{fig:wfs} Isometric view of the  WFS S-H module. At left: currently in SAM. At right: the proposed one for SAMplus. Note that the optics will be greatly simplified owing to  the removal of the Pockels cell.}
\end{figure}

For the WFS optical train, the proposal is to take advantage of the existing design for the lens holders and to make the necessary adjustments according to the new optical design. In this way, the same base plate on the WFS focus stage will be used, without changing the mechanisms. A comparison of the current and new S-H WFS modules can be seen in Fig.~\ref{fig:wfs}. 

\section{Final remarks}
\label{sec:conclusions}
SAMplus is a project to enhance SOAR AO performance at visible wavelengths. The project is based on a more efficient use of the UV laser system available in the SAM with the replacement of more modern adaptive optics components.

Our simulations show a considerable improvement of SAM performance, and a very similar performance for all three different scenarios considered. Therefore, it is preferable the use the DM with 20x20 actuators. The optimal configuration of the WFS subapertures still needs to be determined.

SAMplus operations can be fully integrated with the latest instrumentation. Beyond the obvious case of direct imaging, SAMplus will be integrated with SIFS (SOAR Integral Field Spectrograph), Fabry-Perot, and the future SAMOS spectrograph. 

Extending the possibility of doing AO to the $V$ and $B$ bands will allow investigations in multiple areas of astronomy where angular resolution is critical. These new capabilities of optical AO over a large FoV put SOAR in an unique position as observatory facility, opening new research opportunities.

\acknowledgments 
We thank the SOAR Telescope staff for project assistance. FAPESP grant 2016/16844-1 is acknowledged by DMF; grant 2017/17702-9 by TV; grant 2011/51680-6 by DMF and TV. AM acknowledges support from Serrapilheira grant G-1709-18844. Research developed with help of CCCT HPC-UTFPR.

\bibliography{report} 
\bibliographystyle{spiebib} 

\end{document}